\documentstyle[floats,aps,rotate,epsf,prb]{revtex}

\begin{document}
\twocolumn[\hsize\textwidth\columnwidth\hsize\csname @twocolumnfalse\endcsname

\title{Wave-packet dynamics at the mobility edge in two- and
  three-dimensional systems}

\author{Bodo Huckestein}

\address{Institut f\"ur Theoretische Physik, Universit\"at zu
  K\"oln, D-50937 K\"oln, Germany} 

\author{Rochus Klesse}

\address{Institut f\"ur Theoretische Physik, Universit\"at zu
  K\"oln, D-50937 K\"oln, Germany,\\  and Department of Condensed
  Matter Physics, Weizmann Institute of Science, 76100 Rehovot,
  Israel} 

\date{Received 4 May 1998; revised manuscript received 23 December 1998}

\maketitle

\begin{abstract}
  
  We study the time evolution of wave packets at the mobility edge of
  disordered non-interacting electrons in two and three spatial
  dimensions. The results of numerical calculations are found to agree
  with the predictions of scaling theory. In particular, we find that
  the return probability $P(r=0,t)$ scales like $t^{-D_2/d}$, with the
  generalized dimension of the participation ratio $D_2$. For long
  times and short distances the probability density of the wave packet
  shows power law scaling $P(r,t)\propto t^{-D_2/d}r^{D_2-d}$. The
  numerical calculations were performed on network models defined by a
  unitary time evolution operator providing an efficient model for the
  study of the wave packet dynamics.

\end{abstract}
\vskip2pc

]


The time evolution of wave packets is intimately related to the
transport properties of an electron system. If a wave packet does not
spread unboundedly for long times the system is an insulator. On the
other hand, if a wave packet spreads diffusively for long times the
system has a finite conductivity characteristic of a metal.  In a
seminal paper Anderson observed that disorder can lead to the absence
of diffusion \cite{And58}. With the advent of the scaling theory of
localization \cite{Weg76,AALR79,Weg79} it was realized that the
existence of this Anderson transition between localized and metallic
states depends essentially on the dimensionality and the symmetries of
the system only. In one dimension any amount of disorder leads to
localization of all states while in two dimensions for non-interacting
electrons either a strong magnetic field (as in the quantum Hall
effect \cite{PG87,JVFH94,Huc95r}) or spin-orbit interaction
\cite{HLN80,SZ97,MJH98} are necessary for the existence of extended
states.  Three-dimensional system generally show an Anderson
transition as function of disorder strength and Fermi energy.

At the Anderson transition a system shows a behavior intermediate
between localized and extended. A crucial assumption of the scaling
theory \cite{AALR79} is the finite value of the conductance $G$ of the
system at the transition. The conductance of a $d$-dimensional
hyper-cube of linear dimension $L$ is related to the conductivity
$\sigma$ by Ohm's law $G = \sigma L^{d-2}$. The conductivity at the
transition thus scales like $L^{2-d}$ and so does the diffusion
coefficient $D$ according to the Einstein relation $\sigma = e^2 \rho
D$, with the density of states $\rho$. The Anderson transition is not
reflected in the density of states $\rho$ which is finite through the
transition \cite{Weg79}. Accordingly, the diffusion coefficient is
finite at a two-dimensional transition while there is no diffusive
behavior in three dimensions.

Apart from this scale dependence of the diffusion coefficient a
distinguishing feature of the Anderson transition is the emergence of
multifractal correlations of the wavefunction amplitudes
\cite{Weg76,Aok83,SoEc84,CP86}.  The corresponding density-density correlator
can be expressed in terms of a wave vector- and frequency-dependent
diffusion coefficient $D(q,\omega)$, which in the DC limit becomes the
diffusion coefficient discussed in the previous paragraph. For short
distances and low frequencies, however, this diffusion coefficient
reflects the strong amplitude fluctuations within single
eigenfunctions \cite{Weg76,Cha90}. It should be noted that
multifractal correlations are also present away from the Anderson
transition on scales smaller than the localization length
\cite{FE95a,FE95b}. 

The scaling form of the diffusion coefficient $D(q,\omega)$ has been
discussed previously in the literature
\cite{Cha90,CD88,HHB93,HS94,BHS96}. In this paper we study directly
the time evolution of wave packets. The physical information contained
in both quantities is the same. The advantage of our approach is that
it obviates the need for a diagonalization of the Hamiltonian. Instead
the time evolution operator is applied repeatedly. For systems defined
by a Hamiltonian $H$ this is not necessarily easier to do as the time
evolution operator $U=\exp (-iHt/\hbar)$ is an exponential of the
Hamiltonian and the kinetic and potential terms in the Hamiltonian do
not commute in general complicating the calculation of the
exponential. To circumvent this problem decompositions of the
exponential have been devised \cite{Tro59} and applied to the problem
at hand \cite{KO95,KO96,OK97}. On the other hand, for systems defined
by a unitary operator that can be interpreted as a time evolution
operator the dynamics of wave packets is the most natural and
convenient quantity to study. A class of systems that falls into this
category are the network models introduced by Shapiro \cite{Sha82} and
Chalker and Coddington \cite{CC88}. A network model is characterized
by a unitary operator that describes the scattering of wavefunction
amplitudes at the nodes of a network. Network models have been
constructed and studied for two- and three-dimensional systems with
orthogonal, unitary, symplectic, or chiral symmetry
\cite{KM95,CD95,HK96,KM97,MJH98,FJM99}. The unitary operator of a
network model can be interpreted as a discrete time evolution operator
where a time step corresponds to the typical time of a scattering
event \cite{KlePhD96,KM98}.

The spreading of wave packets has also been discussed in connection
with the nature of the spectrum of the system.
Ketzmerick {\em et al.\/} found that the spreading of wave packets is
determined by the multifractal dimensions $\tilde{D_2}$ and $D_2$ of
the local spectrum and the eigenfunctions, respectively \cite{KKKG97}.
While in general these dimensions are independent quantities
characterizing a system, they are simply related in disordered system,
$D_2 = d \tilde{D_2}$, since the global density of states is scale
invariant and connects length and energy scales via the dimension $d$
\cite{HK96}. In this paper we will restrict our discussion to the case
of the Anderson transition.

After summarizing the scaling behavior of the probability density
of a wave packet, we present the results of numerical simulations for
two- and three-dimensional network models and show that they are in
accordance with the predictions of scaling theory. In particular, we
calculate the return probability to the origin of the wave packet, and
the shape of the wave packet for long times and short distances. We
conclude the paper with a discussion of the results. The scaling forms
for the shape and moments of the wave packet as well as numerical
results for the moments and preliminary results for the shape have
been presented in \cite{HK97a}.


We now summarize the predictions of scaling theory for
characteristic features of a wave packet. For the derivation we refer
the reader to \cite{HK97a} and references therein.

Due to the occurence of multifractal eigenfunction fluctuations at the
localization-delocalization transition, the shape of a wave packet
differs strongly from the diffusive gaussian form. The long time,
short distance behavior, $r^d \ll t/\hbar\rho$ with $\rho$ the density
of states, of $P(r,t)$ is a power law \cite{KKKG97},
\begin{equation}
  \label{prtasympt}
  P(r,t)\propto t^{-D_2/d}r^{D_2-d}.
\end{equation}
The exponent $D_2$ is the generalized dimension of the inverse
participation ratio. For multifractal eigenstates it is smaller than
the space dimension $d$. The term {\em short distance\/} has to be
understood in the sense of $r^d \ll t/\hbar\rho$. The distances on
which Eq.~(\ref{prtasympt}) holds are large compared to any
microscopic scale and correspond for $t\to\infty$ to the limit
$r\to\infty$. For long times Eq.~(\ref{prtasympt}) thus describes the
bulk of the wave packet up to an exponentially small tail.

From eq.~(\ref{prtasympt}) it follows that the return probability to the
starting point of the wave packet $P(r=0,t)$ decays with time as
\begin{equation}
  \label{pt}
  P(r=0,t)\propto t^{-D_2/d}.
\end{equation}

\begin{figure}
  \begin{center}
    \leavevmode
    \epsfysize=8.5cm
    \rotate[r]{\epsffile{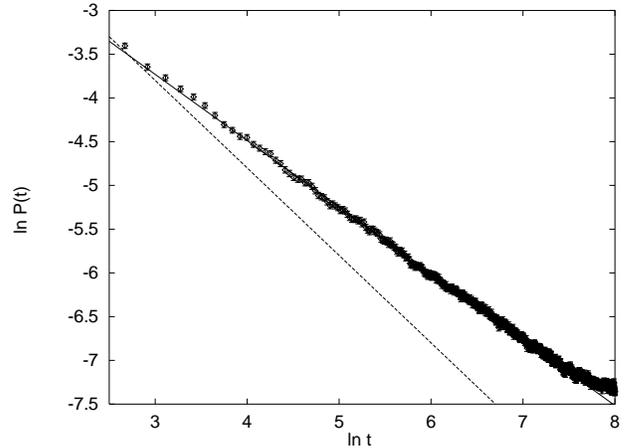}}
  \end{center}
  \caption{Return probability for a wave packet at the quantum Hall
    critical point. The solid line is a best fit to the data with
    slope $D_2/2=0.76\pm0.03$. The dashed line shows the classical
    behavior $p(t)\propto t^{-1}$.}
  \label{fig:2dreturn}
\end{figure}

This is the first quantity that we turn our attention to. It is the
simplest quantity that reflects the multifractality of the eigenstates
at the mobility edge. From Fig.~(\ref{fig:2dreturn}) we get the
exponent $D_2/2$ of the return probability to be $0.76\pm0.03$,
somewhat smaller but still in agreement with the value $0.81\pm0.02$
obtained by Huckestein and Schweitzer for the quantum Hall transition
in a tight-binding model \cite{HS94}. In the present calculation, 400
realizations of the disorder were averaged for a system of
$100\times100$ scatterers.

For a three-dimensional system the exponent $D_2/3$ of the critical
power law differs even more strongly from the diffusive value of
$3/2$. The system that we studied is a stack of quantum Hall networks
coupled in the third dimension \cite{CD95,HK96}. For vanishing
inter-layer coupling the system shows the quantum Hall transition of
the isolated two-dimensional systems, while for finite coupling the
system shows a three-dimensional Anderson transition with localized
and non-localized regimes separated by mobility edges.

\begin{figure}
  \begin{center}
    \leavevmode
    \epsfysize=8.5cm
    \rotate[r]{\epsffile{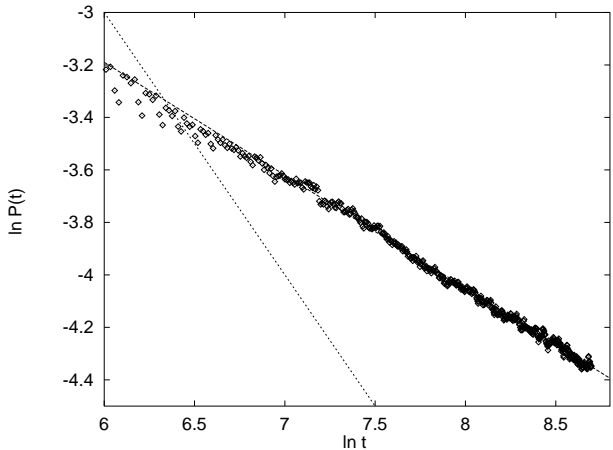}}
  \end{center}
  \caption{Return probability for a wave packet at the Anderson
    transition in $d=3$. The solid line is a best fit to the data with 
    slope $D_2/3=0.43\pm0.04$. The dashed line shows the conventional
    behavior $p(t)\propto t^{-3/2}$.}
  \label{fig:3dreturn}
\end{figure}

The return probability for the three-dimensional network model at the
Anderson transition is shown in Fig.~(\ref{fig:3dreturn}). For the
time interval from 400 to 4900 time steps we obtain a power law with
an exponent $D_2/3=0.43\pm0.04$. For shorter times a crossover to
two-dimensional behavior is observable in our data. Its origin is the
strongly anisotropic character of our system of coupled
two-dimensional networks. The coupling in the third dimension is much
weaker than in the planes so that three-dimensional spreading is
observed only after sufficiently long times. The data presented in
Fig.~(\ref{fig:3dreturn}) were calculated from 200 disorder
realizations of a system of $50^3$ scatterers.

When comparing our result $D_2=1.3\pm0.1$ with values published in the
literature it should be kept in mind that our network model
corresponds to a system with a magnetic field. For time reversal
symmetric systems in the absence of magnetic fields somewhat larger
values have been published previously: $1.7\pm0.3$ \cite{SoEc84},
1.45--1.8 \cite{GS95}, $1.7\pm0.2$ \cite{BHS96}, $1.5\pm0.2$
\cite{OK97}. The values obtained previously in the absence of time
reversal symmetry do not differ significantly from these:
$1.7\pm0.2$ \cite{OK97}, $\approx1.5$ \cite{Ter97}, $1.55\pm 0.15$
\cite{Sch98}. Again our value is somewhat but not significantly
smaller than these values.

\begin{figure}
  \begin{center}
    \leavevmode
    \epsfysize=8.5cm
    \rotate[r]{\epsffile{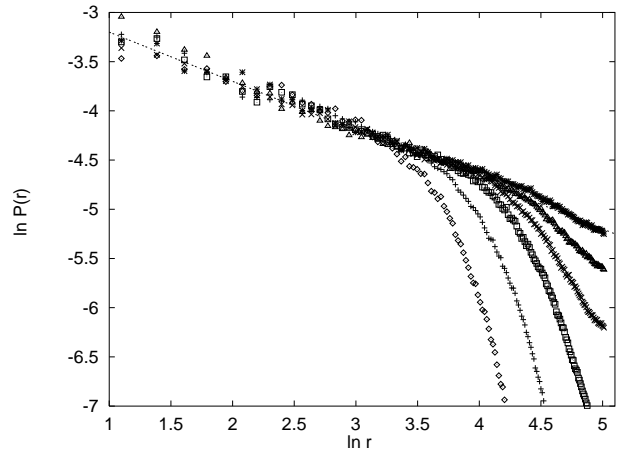}}
  \end{center}
  \caption{Average density of wave packets at the quantum Hall
    critical point. The different curves correspond to different times 
    (1500, 3000, 6000, 9000, 12000, 15000 time-steps). The different
    curves have been rescaled by an amount proportional to
    $t^{0.76}$.}
  \label{fig:2dshape}
\end{figure}

We now show that the bulk of a wave packet has a power law shape. For
the two-dimensional quantum Hall system at the critical point the
average probability density of the wave packet is plotted in
Fig.~(\ref{fig:2dshape}) for different times $t$. The different curves
have been rescaled by a factor of $t^{D_2/2}$ in order to allow for a
better comparison of densities at different times. The straight line
shows the expected power law with exponent $D_2-2=-0.48$ where the
value $D_2$ was taken from the fit to the return probability. The
crossover from power law to exponential shape of the density
distribution is seen to shift with increasing time to larger radii. 6
realizations of the disorder were averaged for a system of
$300\times300$ scatterers.

\begin{figure}
  \begin{center}
    \leavevmode
    \epsfysize=8.5cm
    \rotate[r]{\epsffile{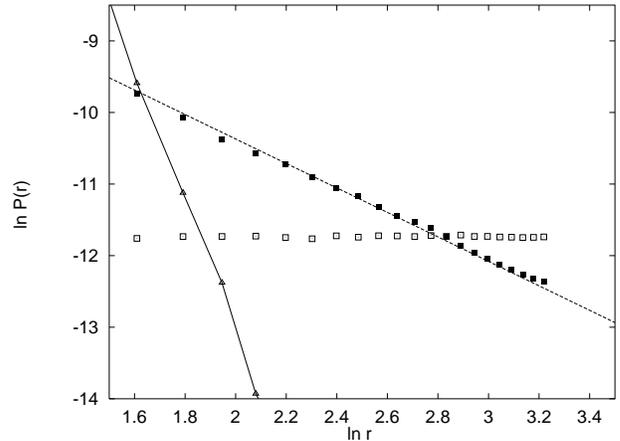}}
  \end{center}
  \caption{Average density of wave packets in three-dimensional
    network models. Data are plotted: in the metallic regime after
    2400 time steps ($\Box$), at the Anderson transition after 
    20000 time steps (\protect\rule{2mm}{2mm}), and in the localized regime
    after 20000 time steps ($\triangle$).} 
  \label{fig:3dshape}
\end{figure}

Figure (\ref{fig:3dshape}) shows the shape of the wave packet in a 
three-dimensional system of $50^3$ scatterers. At the critical point
the power law with an exponent $D_2-3=-1.7$ is observed after 20000
time steps (solid boxes). In the metallic regime the density is
practically constant over the whole system after 2400 time steps (open
boxes). In the localized regime exponential localization is observed
after 20000 time steps (triangles).

To summarize, we have studied the time evolution of wave packets at
the mobility edge in disordered two- and three-dimensional electron
systems. In order to characterize the wave packets we looked at the
return probability and the shape of the density distribution. Due to
the scale invariance of the critical systems these quantities show
power law scaling. The exponents differ from the usual diffusive
situation for two reasons.  On the one hand, at the Anderson
transition the conductance is scale invariant and not the conductivity
or the diffusion coefficient. As a result the spatial dimension $d$
enters where in the diffusive case the exponent $2$ of the diffusion
equation enters. On the other hand, the short distance long time
behavior is governed by multifractal density correlations within
single eigenfunctions leading to the replacement of the spatial
dimension $d$ by the multifractal dimension $D_2$ of the inverse
participation ratio.

We have confirmed the scaling laws and extracted the scaling exponents
by numerical calculations. The calculation were performed for network
models of disordered systems that are defined by a unitary network
operator that serves as a discrete time evolution operator. This
provides a very convenient and efficient way to study the time
evolution of wave packets.

The time dependence of the return probability to the origin shows
power law scaling with the exponent $-D_2/d$ involving the
multifractal exponent $D_2$. Previously, this quantity was studied in
two- and three-dimensional tight-binding models
\cite{HS94,BHS96,OK97}. We obtain values for $D_2$ of $1.52\pm0.06$
and $1.3\pm0.1$ in two and three dimensions, respectively. These
values are somewhat smaller, especially in three dimensions, than
previously published values. At present, it is not clear whether this
difference reflects genuinely different critical behavior in the
systems or different strength of corrections to scaling. It is known
that the two-dimensional network model has very small corrections to
scaling compared to the random Landau matrix model \cite{ECxx,Huc93l}.

For long times $t$ and short distances, $r^d \ll t/\hbar\rho$, the shape
of a wave packet becomes $t^{-D_2/d}r^{D_2-d}$. We observe this
behavior in the two- and three-dimensional network models. At larger
distances the crossover to exponential tails in the density
distribution is observed.

Ketzmerick {\em et al.\/} \cite{KKKG97} obtained under quite general
conditions the expression $P(r,t)\propto t^{-\tilde{D_2}}r^{D_2-d}$
for the shape of wave packets, where $\tilde{D_2}$ is the generalized
dimension of the local density of states. These general results
agree with our results for the Anderson transition since in the
present case the generalized dimensions for the spectrum and the wave
functions are not independent but proportional to each other,
$D_2=d\tilde{D_2}$ \cite{HS94,HK96}. The origin of this simplification
is the scale invariance of the global density of states that allows to
relate energy and length scales via a single relevant energy dependent
length scale $L_\omega\propto \omega^{-1/d}$.


This work was performed within the research program of the
Sonderforschungsbereich 341 of the Deutsche Forschungsgemeinschaft.

\end{document}